\documentclass[a4paper,11pt]{article}
\pdfoutput=1 

\usepackage{jcappub} 

\usepackage[T1]{fontenc} 
\usepackage{subfig}
\usepackage{lscape}
\usepackage{cleveref}

\newcommand{\bea}{\begin{array}}
\newcommand{\ear}{\end{array}}

\newcommand{\bege}{\begin{equation}}
\newcommand{\enge}{\end{equation}}

\newcommand{\beq}{\begin{eqnarray}}\newcommand{\benu}{\begin{enumerate}}\newcommand{\enu}{\end{enumerate}}
\newcommand{\eeq}{\end{eqnarray}}

\newcommand{\me}{\frac{1}{2}}
\newcommand{\noi}{\noindent}

\author[a,1]{Rita C. dos Anjos,\note{Corresponding author.}}
\author[a]{Carlos H. Coimbra-Ara\'ujo,}
\author[b]{Rold\~ao da Rocha}
\author[c]{and Vitor de Souza}

\affiliation[a]{Departamento de Engenharias e Exatas, Universidade Federal do Paran\'a (UFPR),\\
Pioneiro, 2153, 85950-000 Palotina, PR, Brazil.}
\affiliation[b]{Centro de Matem\'atica, Computa\c c\~ao e Cogni\c c\~ao,\\ Universidade Federal do ABC, 09210-580, Santo Andr\'e-SP, Brazil.}
\affiliation[c]{Instituto de F\'isica de S\~ao Carlos, Universidade de S\~ao Paulo, \\ Av. Trabalhador S\~ao-carlense 400, S\~ao Carlos, Brazil}

\emailAdd{ritacassia@ufpr.br}
\emailAdd{carlos.coimbra@ufpr.br}
\emailAdd{roldao.rocha@ufabc.edu.br}
\emailAdd{vitor@ifsc.usp.br}

\title{Ultra high energy cosmic rays and possible signature of black strings}

\abstract{Ultra high energy cosmic rays (UHECRs) probably originate in
  extreme conditions in which extra dimension effects might be
  important. In this paper we calculate the correction in black hole
  accretion mechanisms due to extra dimension effects in the static and
  rotating cases. A parametrization of the external Kerr horizons in
  both cases is presented and analysed. We use previous calculations of
  upper limits on the UHECR flux to set limits on the UHECR production
  efficiency of nine sources. The upper limit on the UHECR luminosity calculation is based on GeV-TeV gamma-ray measurements. The total luminosity due to the accretion mechanism is compared to the upper limit on UHECRs. The dependence of the UHECR production efficiency upper limit on black hole mass is also presented and discussed.}

\begin{document}
\maketitle
\flushbottom


\section{Introduction}

The possible presence of extra dimensions in the universe is one of the
most astounding features of string theory. Despite the strong theory
formalism, extra dimensions still remain inaccessible to experiments~\cite{experiments1,experiments2,experiments3,experiments4,experiments5,experiments6}. Since the presence of ten or more spacetime dimensions is one of the central conditions of string theory and M theory~\cite{gr,zwi,zwi1,zwi2}, it is not unrealistic to say that experimental observation or constraints on the extra dimensions properties would be a major advance in science.

 Randall-Sundrum (RS) models are capable of explaining the hierarchy
 problem~\cite{Randall1,Randall2} and Kaluza-Klein (KK) non compactified
 theories~\cite{Others}. The extra dimension implies deviations from
 Newton's law of gravity at sub millimetric scales, where objects may be
 indeed be gravitating in higher dimensions \cite{Randall1,Randall2}. 
This main assumption predicts deviations from $4D$ general relativity in
the dynamics of massive astrophysical objects, e.g. black holes
(BH). The generalization of a BH in the brane world models usually
involve black strings. Within this  framework, the total luminosity of
Active Galactic Nuclei (AGNs) and radio galaxies have been calculated in
references~\cite{roldao1,roldao2,roldao3} for an effective extra dimension size. The general conclusion is that the total luminosity generated by the accretion mechanism is reduced due to possible extra dimension corrections predicted by RS and KK models.

Ultra high energy cosmic rays (UHECRs - $E > 10^{18}$ eV) are produced
in the same environment in which extra dimension effects might be
important. The recent results of the Pierre Auger Observatory limiting
drastically the flux of gamma rays~\cite{PAC} and neutrinos~\cite{PAC1}
with energy above $10^{18}$ eV have minimised the importance of exotic
non-acceleration production of UHECR known as top-down
mechanisms. Moreover,the study of the arrival direction of the events
measured by the Pierre Auger Observatory with energy above 58 EeV cannot
reject, on a 2 sigma level, a correlation with local mass and a source distribution compatible with the known AGN spacial distribution~\cite{PA}.

From the theory point of view, the potential of a source to accelerate
particles up to the extreme energies is summarized in the Hillas
plot~\cite{hillas84}. Following his argument, a short list of the most
probably acceleration sites can be selected by requiring that during
acceleration the particles must be confined in the source, in other
words, the size of the source must be larger than the Larmor radius of
the particle in the magnetic field present in the acceleration
region. Further arguments have been developed in order to pinpoint
source candidates concerning the acceleration time scale~\cite{norman,henri} and jet power~\cite{lovelace}. The convergence of such studies means that the leading contenders for accelerating particles to the highest energies are the most powerful radio galaxies, AGNs, gamma-ray bursts, fast spinning newborn pulsars and interacting galaxies. For details about the acceleration see references~\cite{takahara,rachen,nagano1,bluemer,letessier,olinto10,lemoine12,watson13}.

The accretion mechanism in BH is known to be the main source of energy which powers all sorts of emission. The luminosity in each channel (X) is usually written as a fraction of the total power of the accretion mechanism ($L_{acc})$), i.e. $L_X = \eta_X L_{acc}$. The linear correlation between the accretion mechanism and power of the compact radio core was studied in reference~\cite{falcke1995}. References~\cite{saikia,miller} suggests that $\eta_{X-ray} \sim 10^{-5}$. A recent publication~\cite{Dutan2015} has investigated the acceleration of UHECRs in low-luminosity AGNs suggesting that $\eta_{CR} \sim 1/3$. However the role of UHECR in the energy balance of the accretion mechanism is yet to be carefully studied.

Recently, a method has been developed to calculate upper limits of UHECR
luminosities of individual sources~\cite{supa}. The method is based on
the measurement of an upper limit of the integral flux of GeV-TeV gamma
rays. In reference~\cite{vitor}, this method was used to set the upper
limit on the UHECR luminosity of twenty nine sources. The calculated
upper limits on the proton and the total UHECR luminosity are of the order of $\sim 10^{45}$~erg/s.

In the present paper, we calculate the luminosity of nine of these sources taking into account deviations from general relativity due to RS bulk  and compare the results. Considering the extra dimension corrections, the predicted total luminosity goes from the fundamental Eddington limit ($\sim 10^{47}$~erg/s) to $\sim 10^{45}$~erg/s which is the same order of the UHECR luminosity calculated in reference~\cite{vitor}. We studied the dependence of the black string radius as a function of the total mass and a parametrization is proposed.

The bolometric luminosities measured for AGNs and radio galaxies are
roughly $\sim 10^{42}-10^{43}$ erg/s~\cite{marek} which is orders of
magnitude below the values predicted by accretion mechanisms ($\sim
10^{45}-10^{47}$ erg/s) and therefore do not to constrain the models. On
the other hand, we show here that the upper limits of the UHECR are of
the same order as the flux predicted by accretion mechanisms with extra dimension effects in AGNs and radio galaxies. For seven out of nine sources studied here, we manage to set an upper limit on $\eta_{CR/pr}$. Extreme cases of pure proton and pure iron nuclei emission are considered. The case of pure iron nuclei limits the total UHECR flux as explained in reference~\cite{vitor}.

We also investigate the dependence of $\eta_{pr}$ on the mass of the BH with and without extra dimension corrections. We show a hint of decreasing $\eta_{CR}$ for increasing BH mass.

The article is organized as follows. In Section~\ref{sec:bs} we review
the calculation of the BH radius in the static and rotating case with and without extra dimension corrections. A simple parametrization of the solutions is presented. In this section, the accretion mechanism powering the BH luminosity is also summarized. In Section~\ref{sec:cr} the calculation of the upper limit on the UHECR luminosity is presented. Section~\ref{sec:conclusion} concludes this study.

\section{Black hole accretion with extra dimension corrections}
\label{sec:bs}

The approach in this section is to summarize  corrections in black hole horizons due to brane-world effects, consisting of the black string framework. Static and rotating black holes are calculated both for Randall-Sundrum (RS) and Casadio-Fabbri-Mazzacurati metrics in references~\cite{roldao1} and~\cite{roldao2}, respectively. In this paper, we use the RS formalism in which volcano barrier-shaped effective potentials for gravitons are induced around the brane~\cite{Likken}. Such bulk modes introduce small corrections at short distances. Compact dimensions do not affect the localization of matter fields, however true localization is achieved merely for massless fields~\cite{Gregory}. In the massive case, the bound state is metastable and may leak into the extra space.

The RS metric is in general expressed as

\begin{equation}
^{(5)}ds^2 = e^{-2k|y|}g_{\mu\nu}dx^{\mu}dx^{\nu} + dy^2,
\end{equation}

\noindent
where $k^2 = 3/2\ell^2$ and $\ell$ denotes the Ad$S_{5}$ bulk curvature radius. Moreover, the term $e^{-2k|y|}$ is the so called \emph{the warp factor} \cite{Randall1,Randall2,Maartens}, which confines gravity in the brane at low energies \cite{Randall1,Randall2,experiments5,Maartens}. 
It is worth to mention that in the original RS formalism $g_{\mu\nu}$ is the Minkowski metric \cite{Randall1,Randall2,experiments5}, however a more general setup will be considered here. The Gaussian coordinate $y$ denotes the normal direction out of the brane into the bulk.

In a braneworld framework embedded in AdS$_5$ bulk, the Einstein field equations are expressed as

\begin{eqnarray}\label{123}
&&G_{\mu\nu} = -\frac{1}{2}{\Lambda}_5g_{\mu\nu}+ \frac{1}{4}\kappa_5^4\left[\frac{1}{2}g_{\mu\nu}(T^2 - T_{\rho\sigma}T^{\rho\sigma})- T^{\;\;\rho} _{\nu}T_{\mu \rho}+TT_{\mu\nu}\right] - E_{\mu\nu},\nonumber
\end{eqnarray}

\noindent
where $\Lambda_5$ is  
the 5D cosmological constant, $T = T_\rho^{\;\;\rho}$ denotes the momentum-energy tensor trace, $\kappa_5^2=8\pi G_5$ (with $G_5$ the
5D  Newton constant) and  $E_{\mu\nu}$ denotes the  components of the
electric part of the Weyl tensor, that describes high
energy effects originating Kaluza-Klein (KK) modes and  information 
regarding collapse processes of BHs as well. In particular, the gravitational collapse induces  deviations from the 4D general
relativity \cite{experiments5}.  RS KK  modes provide a deviation of the Newtonian potential as \cite{Randall1,Randall2}
\begin{equation}\label{potential}
V(r) = \frac{GM}{c^2r}\left(1 + \frac{2\ell^2}{3r^2} + \mathcal{O}\left(\frac{\ell^4}{r^4}\right)\right).
\end{equation}
\noindent

Some useful equations can be derived forthwith from Bianchi identities $\nabla_\mu G^{\mu\nu}=0$ and vacuum on the brane $T_{\mu\nu}=0$. For instance we have $\nabla_\nu E^{\mu\nu}=0$ (see e.g.~\cite{roldao1,Maartens,Shiromizu}). Moreover a
perturbative procedure is then brought into play, to study the bulk metric near the brane, which in particular further inherits the black string 
warped horizon along the extra dimension as well. 
It is accomplished by a Taylor expansion to probe properties of a static BH on the brane~\cite{Da}. Such expansion, for a vacuum brane metric, up to terms of order ${\mathcal O}(y^5)$, reads

\begin{eqnarray}\label{metrica}
&&\negthickspace g_{\mu\nu}(x,y) = g_{\mu\nu}(x,0) - E_{\mu\nu}(x,0)\frac{y^2}{2!} - \frac{12}{\ell}E_{\mu\nu}(x,0)\frac{|y|^3}{3!}\nonumber\\
&&\qquad\qquad+2\left[6E_{\mu}^{\; \alpha}E_{\alpha\nu}+ 2E^{\alpha\beta}R_{\mu\alpha\nu\beta}+\left({\Box} - \frac{32}{\ell^2}\right)E_{\mu\nu} \right]_{y=0}\negthickspace \frac{y^4}{4!}+\cdots
\nonumber
\end{eqnarray}

\noindent where $\Box$ denotes the standard  d'Alembertian, showing that the propagating effect of $5D$ gravity is  merely evinced at the fourth order. A more general expression was derived in \cite{hoff}.

\subsection{Static black strings}

For a static spherical metric on the brane 

\begin{equation}\label{124} g_{\mu\nu}dx^{\mu}dx^{\nu} = - F(r)dt^2 + \frac{dr^2}{H(r)} + r^2d\Omega^2,
\end{equation}

\noindent where $d\Omega^2$ denotes the spherical 2-volume element in the 3-brane. Moreover, the electric part of the Weyl tensor on the brane has the following components:
\begin{eqnarray}
E_{00} &=& \frac{F}{r}\left(\frac{H-1}{r}+H'\right),\quad
E_{\theta\theta} =  H -1+\frac{rH}{2}\left(\frac{H'}{H}+\frac{F'}{F}\right),\nonumber\\
E_{rr} &=& \frac{1}{r}\left(\frac{F'}{FH} + \frac{ H^{-1}-1}{r}\right)
\end{eqnarray}

\noindent By denoting $\psi(r)$ a deviation from the usual  Schwarzschild metric term  \cite{Maartens,rs05,rs06,rs07,rs01,rs02,rs03,Gian}
\begin{equation}\label{h}
H(r) = 1 - \frac{2GM}{c^2r} + \psi(r),
\end{equation}

\noindent
where $M$ is usually identified to the effective mass and, given the expansion of the warped horizon area (determined by $g_{\theta\theta}$),

\begin{eqnarray}\label{gtheta}
g_{\theta\theta}(r,y) &=& r^2  - \psi'\left(1 + \frac{2}{\ell}|y|\right)y^2\nonumber\\ +\negthickspace \negthickspace &&\negthickspace\negthickspace\left[\psi' + \frac{1}{2}(1 + \psi')(r\psi' - \psi)'\right]\frac{y^4}{6r^2} + \cdots,
\end{eqnarray}

\noindent
the function  $\psi$ can be shown to determine the modification of the warped horizon areal radius  along the extra dimension \cite{Maartens}.  For a black hole with horizon $r \gg \ell$, eq.(\ref{potential}) yields

\begin{equation}\label{psi}
\psi(r) \approx -\frac{4GM}{3c^2r}\left(\frac{\ell^2}{r^2}\right).
\end{equation}

\noindent
The corrected horizon $R_{Sbrane}$ for the static BH including extra dimension corrections comes by fixing $H(r) = 0$, i.e. by solving equation $1 - \frac{2GM}{c^2R_{Sbrane}} + \psi(R_{Sbrane})=0$ in  eq.(\ref{h}), resulting in \cite{roldao1,roldao2,roldao3}

\begin{equation}\label{111}
R_{{\rm Sbrane}} = \frac{2GM}{3c^2}+ (g_2 - \sqrt{g_1})^{1/3} + (g_2 + \sqrt{g_1})^{1/3} ,\end{equation}

\noindent where

\begin{eqnarray}
 g_1&=&\frac{4G^2M^2\ell^2}{9c^4}\left(\frac{8G^2M^2}{9c^4}+\ell^2\right)\,\label{rs4}\\
 g_2 &=& \frac{2GM}{3c^2}\left(\frac{4G^2M^2}{9c^4}+\ell^2\right),\label{rs3}
\end{eqnarray}
 
\subsection{Rotating black strings}
\label{sec:rotation}

The spherical rotating BH Kerr metric is represented by the following metric tensor:

\bege\label{eve} g_{\mu\nu}^{\rm Kerr} = \begin{pmatrix}
\omega^2\tau^2 - \alpha^2&0&0&-\omega\tau^2\\
0&{\rho^2}/{\Delta}&0&0\\
0&0&\rho^2&0\\
-\omega\tau^2&0&0&\tau^2
\end{pmatrix}\enge\noi where the BH presents angular momentum $J$ and mass parameter  $M$, and 

\beq
\rho^2 &=& \frac{a^2}{c^2}\cos^2\theta+r^2\,,\qquad
\Delta = \frac{a^2}{c^2}  - \frac{2GM}{c^2}r + r^2,\qquad 
\tau = \frac{\Sigma}{\rho}\sin\theta,\nonumber\\
\alpha &=& \sqrt{1 - \frac{2GM}{c^2r}},\qquad \omega = \frac{2aGMr}{c^3\Sigma^2},\qquad 
\Sigma^2 = \left(\frac{a^2}{c^2}+r^2\right)^2 - \frac{a^2}{c^2}\Delta\sin^2\theta\,.
\eeq

The rotation parameter $a$ is defined by $a = \frac{J}{Mc} \sim \frac{GM}{2c}$. The associated diagonal metric reads

\beq\label{kerr}
g^{\rm Kerr} &=& g_{\mu\nu}^{\rm Kerr}dx^\mu dx^\nu\nonumber\\
 &=& \lambda_0 {dt^{\prime}}^2 + \frac{\Delta}{\rho^2}{dr^2} + \rho^2 d\theta^2 + \lambda_3d{\phi^\prime}^2,\label{metricap}
\eeq

\noi where

\beq
\lambda_{0,3} &=& \me\left[(\omega^2\tau^2 - \alpha^2+\tau^2)\right. \nonumber\\
&\pm& \left.\left((\omega^2\tau^2 - \alpha^2+\tau^2 )^2 - 4(\omega\tau^2+(\omega^2\tau^2+\tau^2)^2)\right)^{1/2}\right].
\eeq

\noi The inner and outer Kerr horizons ($R_{\pm \rm Kerr}$) are achieved when the coefficient ${\frac{\rho^2}{\Delta}}$  in the denominator  of $dr^2$ in eq.(\ref{kerr}) is equal to zero. Hence it reads 

\bege\label{raio}
R_{\pm \rm Kerr}= \frac{GM}{c^2}\pm\sqrt{\left ( \frac{GM}{c^2} \right )^2 - \frac{a^2}{c^2}} \; .
\enge

\noi Now, using the approximation for $a$ we can solve for $R_{+ \rm Kerr}$ and show a straightforward dependency with $M / M_{\odot}$

\bege\label{raio2}
R_{+ \rm Kerr}= \frac{GM}{c^2} + \frac{\sqrt{3}}{2} \frac{GM}{c^2} \sim 1.38\times10^{-27} M \; {\rm [m/kg]} \sim 2.73 \frac{M}{M_{\odot}} \; {\rm [m]}\; .
\enge

In order to obtain the correction of the Kerr horizons in eq.(\ref{raio}) due to braneworld effects ($R_{\rm Kerr brane}$), we follow the idea presented in eq.(\ref{h}). Defining $\xi(r)$ as being the deviation from a Kerr standard metric for ${\frac{\rho^2}{\Delta}}$ (namely the term in the denominator of $dr^2$ in eq.(\ref{metricap})), we provide the \emph{ansatz}

\bege\label{133}
\xi(r) = -\frac{4GM\ell^2}{3r(a^2+c^2r^2)}
\enge

\noi such that in the static limit where $a\rightarrow 0$ the correction induced in the gravitational potential is consistent with eq.(\ref{potential}), and to eq.(\ref{psi}), accordingly. The corrections in the Kerr event horizons, as calculated in \cite{roldao1},  are achieved through the deformed Kerr metric, as

\bege
 \frac{\rho^2}{\Delta} + \xi(R_{\rm Kerr brane}) = 0.
\enge

\noi When the above expression is expanded and using eq.(\ref{133}) it yields 

\beq\label{cin}
&&c^2R_{\rm Kerr brane}^5 - 2GMR_{\rm Kerr brane}^4 + 2a^2R_{\rm Kerr brane}^3 \nonumber\\&&- GM\left(2\frac{a^2}{c^2} + \frac43\ell^2\right)
R_{\rm Kerr brane}^2 + \frac{a^4}{c^2}R_{\rm Kerr brane}\nonumber\\&& - \frac43 GM\ell^2 \frac{a^2}{c^2}\cos^2\theta = 0\,.
\eeq

\begin{figure}
  \centering
  \includegraphics[width=0.9\textwidth]{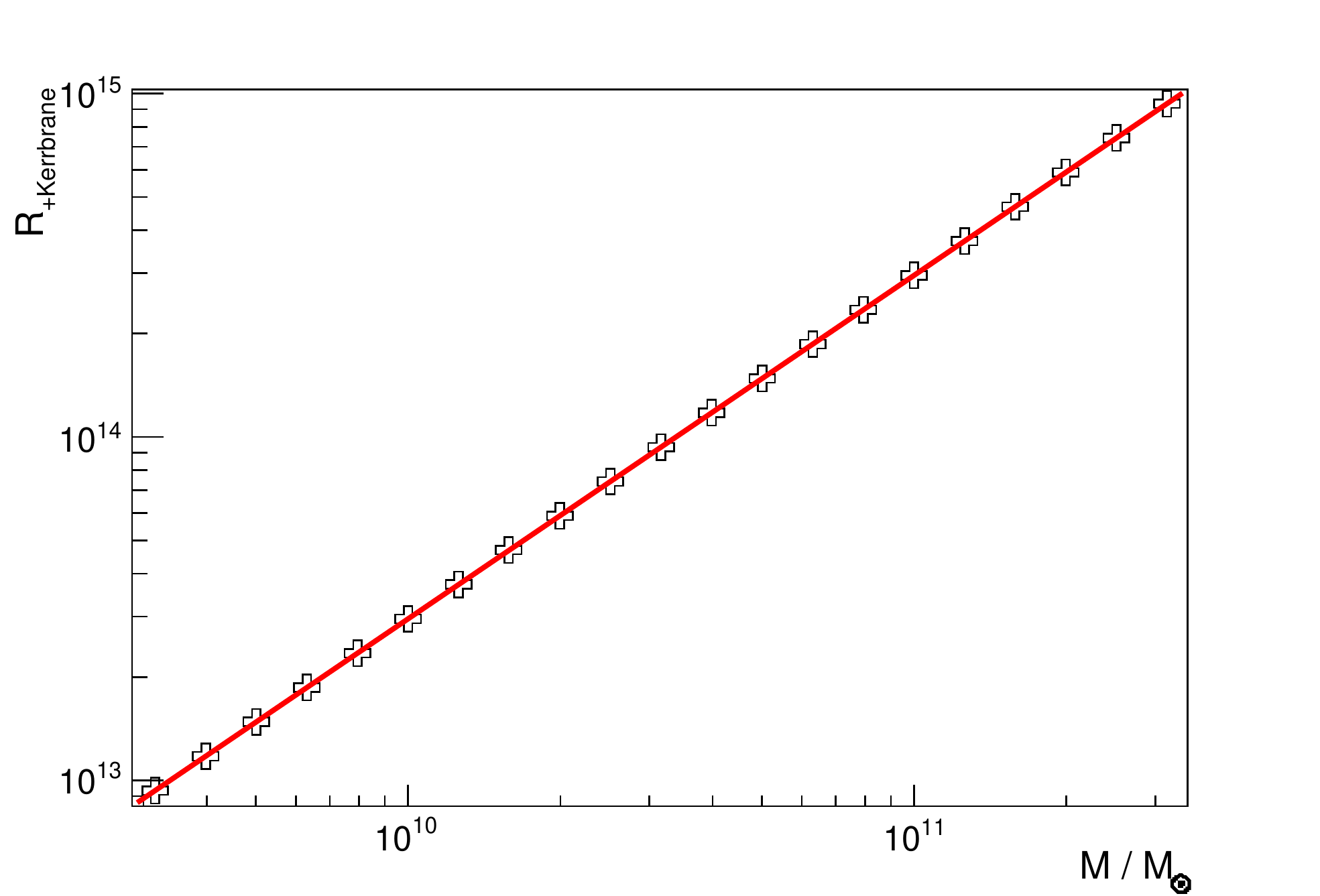}
  \caption{$R_{{+\rm Kerr brane}}$ as a function of $M/M_\odot$. Crosses show the solution of equation~\ref{cin} for$M/M_\odot = [9.5,11.5]$ in steps of 0.1. Red line is a linear fit to the points $R_{+ \rm Kerr brane} = a + b\times M/M\odot$ with $a = 5\times10^{6} \pm 5\times10^{7}$ m and $b = 2954.12 \pm 0.00047 $ m.}
  \label{fig:rkb:mass}
\end{figure}

\noi See reference~\cite{roldao1} for a detailed description about those solutions and other possible groups of $\xi(R_{\mathrm{Kerr brane}})$ \emph{ans\"atze}. 
We have numerically solved equation~(\ref{cin}) to find
$R_{+ \rm Kerr brane}$. Figure~\ref{fig:rkb:mass} shows the dependency of $R_{+ \rm Kerr brane}$ with the black hole mass. A straightforward linear relation holds, leading to $R_{+ \rm Kerr brane} = a + b\times M/M_\odot$ with $a = 5\times10^{6} \pm 5\times10^{7}$ m and $b = 2954.12 \pm 0.00047 $ m.

\subsection{Luminosity of black strings}

The luminosity of AGNs are produced essentially by the accretion mechanism, i.e., AGN activity is related to the growth, via accretion, of central supermassive black holes (SMBHs). The accretion efficiency ($\epsilon$), from a simple accretion model, is given by (see, e.g., \cite{shapiro})

\begin{equation}\label{eta1}
\epsilon = \frac{GM}{6c^2R}
\end{equation}

\noindent
where $R = R_{{\rm Sbrane}}, R_{+\rm Kerr}$ or $R_{{+\rm Kerr brane}}$ are the horizons for a static BH with extra dimension correction, a rotating BH without extra dimension corrections, and a rotating BH with extra dimension corrections, respectively. The luminosity $L_{acc}$ of AGNs due to BH accretion is given by

\begin{equation}\label{dl}
L_{acc}^{Theory} = \epsilon  \dot{M}c^2 = \frac{GM\dot{M}}{6R} \; ,
\end{equation}

\noindent
where $R = R_{{\rm Sbrane}}, R_{+\rm Kerr}$ or $R_{{+\rm Kerr brane}}$. The varying mass term $\dot{M}$ denotes the accretion rate and depends on some specific model of accretion. Several accretion rates are proposed in the literature, for comparison we use here the Bondi mechanism $\dot{M} = \pi \lambda c_{S} \rho_B r_B$ \cite{bondi}, where $\lambda$ takes the value of 0.25 for an adiabatic index 5/3, $c_S$ is the sound speed in the medium, $r_B$ denotes the Bondi accretion radius and $\rho_B$ stands for the gas density at that radius. For simplicity we use $\dot{M} = 10^{12} M/M\odot {\rm [kg.s^{-1}]}$.

Considering a fraction $\eta_{CR}$ of the total luminosity goes into UHECR we write

\begin{equation}\label{lcr}
L_{CR}^{Theory}=\eta_{CR} L_{acc}^{Theory} \; .
\end{equation}

\section{Upper limit on the UHECR emission efficiency}
\label{sec:cr}

In reference~\cite{supa} a method was developed to set upper limits on the UHECR luminosity of individual sources, based on gamma-ray measurements. We summarize this method below.

UHECRs are produced in extragalactic sources and propagate in the background radiation fields. The particles interact with background photons and start a cascade process. Successive interactions of the particles produce large numbers of electrons, positrons, neutrinos and photons. The dependence of the mean free path of the gamma-gamma interaction with energy results in a large abundance of GeV-TeV photons produced in the cascade. The relevant processes and the details of the particle production are explained by references~\cite{bib:pp:1,bib:pp:2,bib:pp:3,bib:DeAngelies:13}.

The method connects a measured upper limit on the integral flux of GeV-TeV gamma-rays and the UHECR cosmic-ray luminosity ($L_{CR}^{UL}$) of a source by the equation:

\begin{equation}
L_{CR}^{UL} = \frac{4\pi D^{2}_s(1+z_s)\langle E \rangle_{0}}{ \ K_{\gamma} {\mathop{\displaystyle
\int_{E_{th}}^{\infty} dE_\gamma\ P_{\gamma}(E_{\gamma})}}}\ I_{\gamma}^{UL}(> E^{th}_{\gamma}),
\label{eq:CRUL}
\end{equation}

\noindent
where $I_{\gamma}^{UL}(> E^{th}_{\gamma})$ is the upper limit on the integral gamma-ray flux for a given confidence level and energy threshold, $K_{\gamma}$ is the number of gamma-rays generated from the cosmic-ray particles, $P_{\gamma}(E_{\gamma})$ is the energy distribution of the gamma-rays arriving on Earth, $E_{\gamma}$ is the energy of gamma-rays, $\langle E \rangle_{0}$ is the mean energy, $D_s$ is the comoving distance and $z_s$ is the redshift of the source.

The procedure to apply this method to calculate $L_{CR}^{UL}$ for a
given source is as follows. $D_s$ and $z_s$ must be know for the source
from astronomical measurements. $I_{\gamma}^{UL}(> E^{th}_{\gamma})$ is
measured by gamma-ray observatories for each sources. $E^{th}$ is given
by the gamma-ray observatory and by the data analysis. $K_{\gamma}$ and
$P_{\gamma}(E_{\gamma})$ are calculated given a primary UHECR energy
spectrum. The type of the primary UHECR is needed for the calculation of $K_{\gamma}$ and $P_{\gamma}(E_{\gamma})$. We have done calculations for proton and iron nucleus primaries. Since iron nuclei are the particles to produce the largest numbers of GeV-TeV gamma rays, the limit calculated with iron nuclei simulations imposes an upper limit on the total UHECR luminosity. For details about the calculation see references~\cite{supa,vitor}.

Twenty nine sources have been studied in references~\cite{vitor} and upper limits on the proton or total UHECR luminosity were set. The sources have been measured by the FERMI-LAT~\cite{fermi}, H.E.S.S~\cite{hess}, MAGIC~\cite{magic} and VERITAS~\cite{veritas} experiments. We found in the literature the mass estimative of nine of these sources, which are used here. Other UHECR upper limits were calculated in \cite{unger} and \cite{waxman}. 

The data of each source is shown in tables~\ref{tab:1} and~\ref{tab:2}. Table~\ref{tab:1} summarizes the results for sources that only a upper limit on the proton luminosity was set and table~\ref{tab:2} shows the same results for sources that the total UHECR luminosity was set.  The values of $L_{acc}^{Theory}$ for $R = R_{{\rm Sbrane}}, R_{+\rm Kerr}$ or $R_{{+\rm Kerr brane}}$ are shown. We use the upper limits on the UHECR ($L_{CR}^{UL}$) to set upper limits on $\eta_{CR}$: $\eta_{CR}^{UL} = L_{CR}^{UL}/L_{acc}^{Theory}$. These limits are valid only if $L_{CR}^{UL} < L_{acc}^{Theory}$. If only a upper limit of the proton UHECR was set, we calculate the corresponding $\eta_{pr}^{UL}$. We show the corresponding $\eta_{pr}^{UL}$ for $R = R_{+ \rm Kerr}$ and $R_{+ \rm Kerr brane}$.

\section{Concluding Remarks}
\label{sec:conclusion}

We compare in tables~\ref{tab:1} and~\ref{tab:2} the luminosity of the BH with and without extra dimension corrections. The luminosity decreases significantly for some cases when extra dimension effects are taken into consideration. On average \mbox{$<L_{+Kerr brane}> \; =  \; 1.08 <L_{+Kerr}>$}.

\begin{figure}
  \centering
  \includegraphics[angle=90,width=0.9\textwidth]{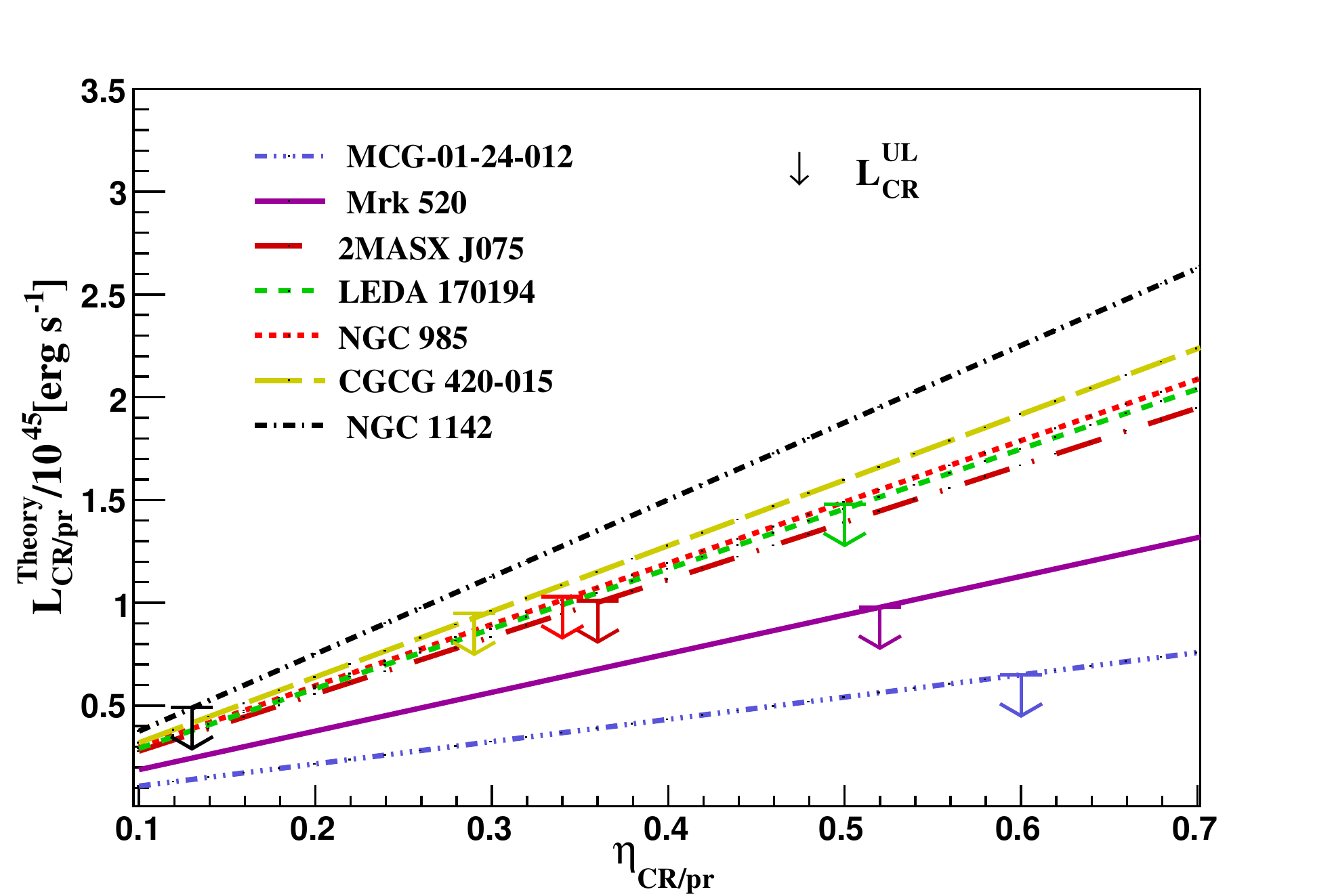}
  \caption{The plots show $L_{CR}^{Theory}$ as a function of the
    parameter $\eta_{CR}$. The arrows indicate the values of the
    luminosity of ultra-high-energy cosmic rays $L_{CR}^{UL}$ calculated using the integral flux of GeV-TeV gamma-rays \cite{vitor}.}
  \label{fig:luminosity}
\end{figure}

Figure~\ref{fig:luminosity} shows $L_{CR}^{Theory}$ as a function of the $\eta_{CR/pr}$ \footnote{CR for total luminosity or pr for proton only.}. The arrows indicate $L_{CR/pr}^{UL}$ calculated using the integral flux of GeV-TeV gamma-rays~\cite{vitor} and therefore sets $\eta_{CR/pr}^{UL}$.

\begin{figure}
  \centering
  \includegraphics[width=0.9\textwidth]{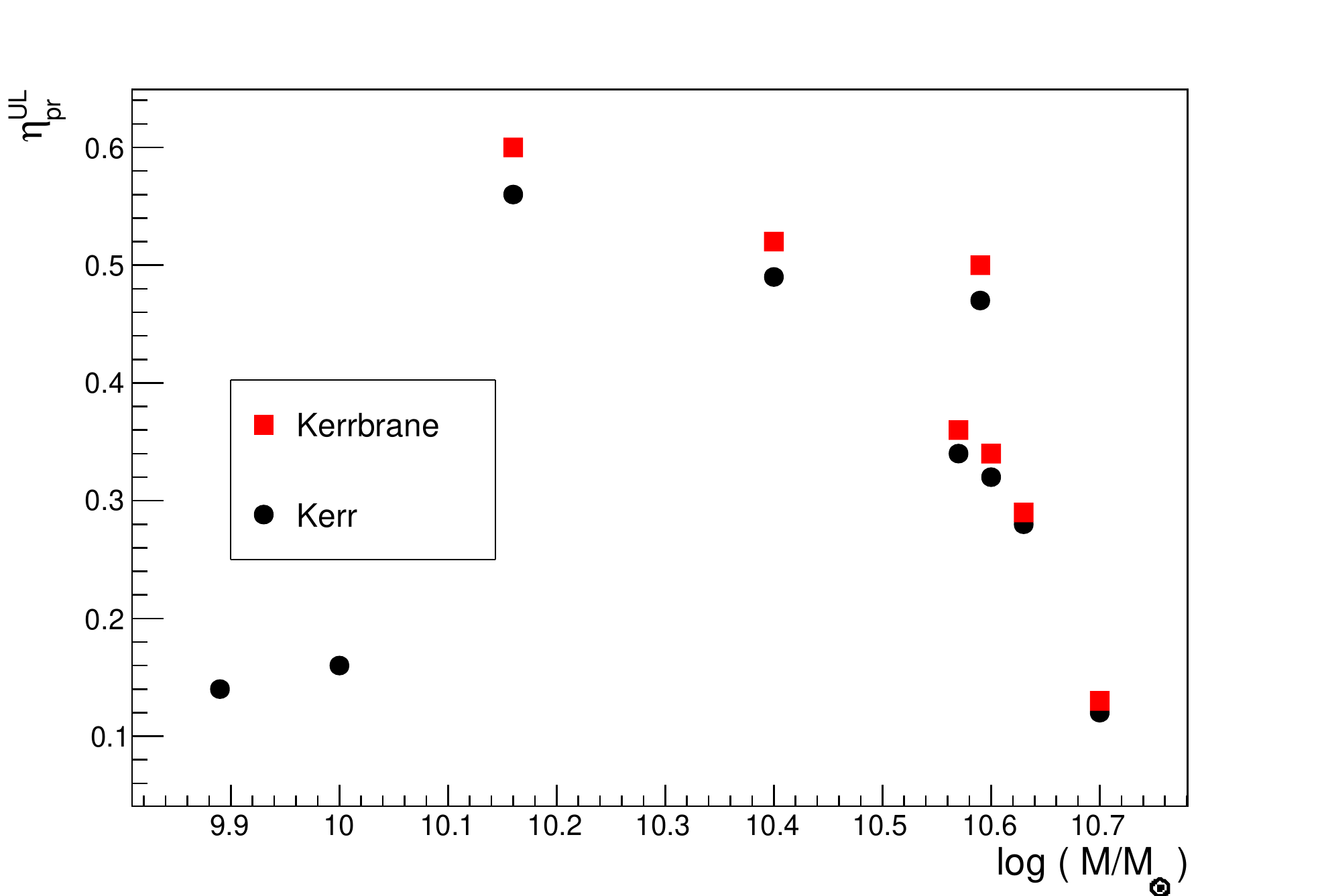}
  \caption{$\eta_{pr}^{UL}$ as a function of $log ( M/M_\odot )$. The graphics show the results the calculation of $\eta_{pr}^{UL}$ for the standant Kerr BH (Kerr - black circles) and for Kerr BH with extra dimension correction (Kerr brane - red squares).}
  \label{fig:mass:eta}
\end{figure}

The calculated values of $\eta_{pr}^{UL}$ varies from 0.12 (0.13) to 0.69 (0.73) with average 0.36 (0.43) and standard deviation 0.18 (0.18) for the Kerr (Kerr brane) cases. These values, despite being only upper limits are compatible to the estimatives done in reference~\cite{falcke1995,Dutan2015}. Figure~\ref{fig:mass:eta} shows that $\eta_{pr}^{UL}$ decreases with increasing BH mass for $R = R_{{+\rm Kerr brane}}$. If this relation reveals a fundamental behaviour of the accretion mechanism or a feature of the methods developed in this paper is a subject to be studied. If $R = R_{{+\rm Kerr}}$ the picture is not so clear because of the two sources for which an upper limit on $\eta_{pr}^{UL}$ was set only for $R = R_{{+\rm Kerr}}$.

\begin{figure}
  \centering
  \subfloat[$L_{acc}^{Theory}$ x $\ell$ study.]{\includegraphics[angle=90,width=0.8\textwidth]{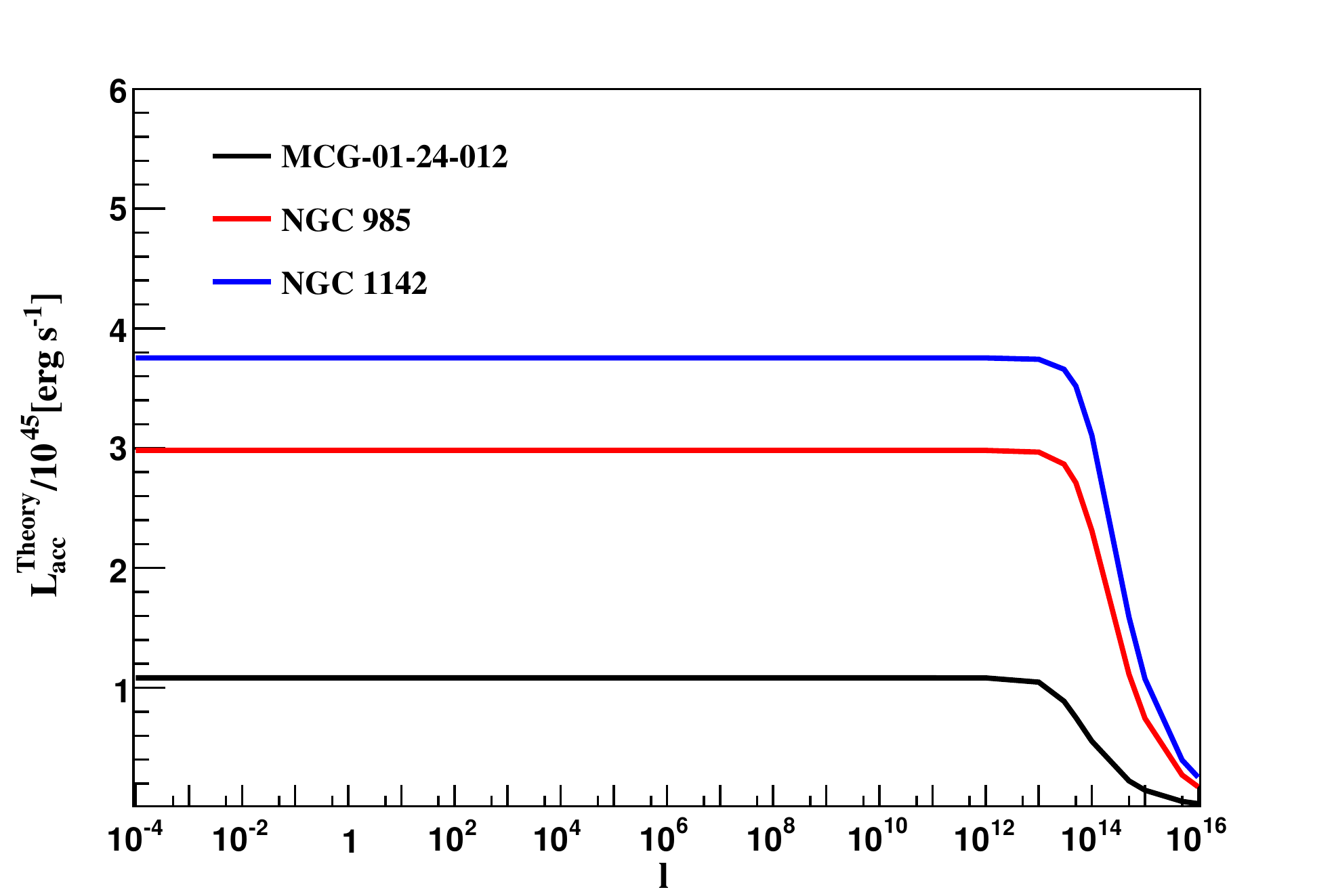}}\\
  \subfloat[$\eta_{pr}^{UL}$ x $\ell$ study.]{\includegraphics[angle=90,width=0.8\textwidth]{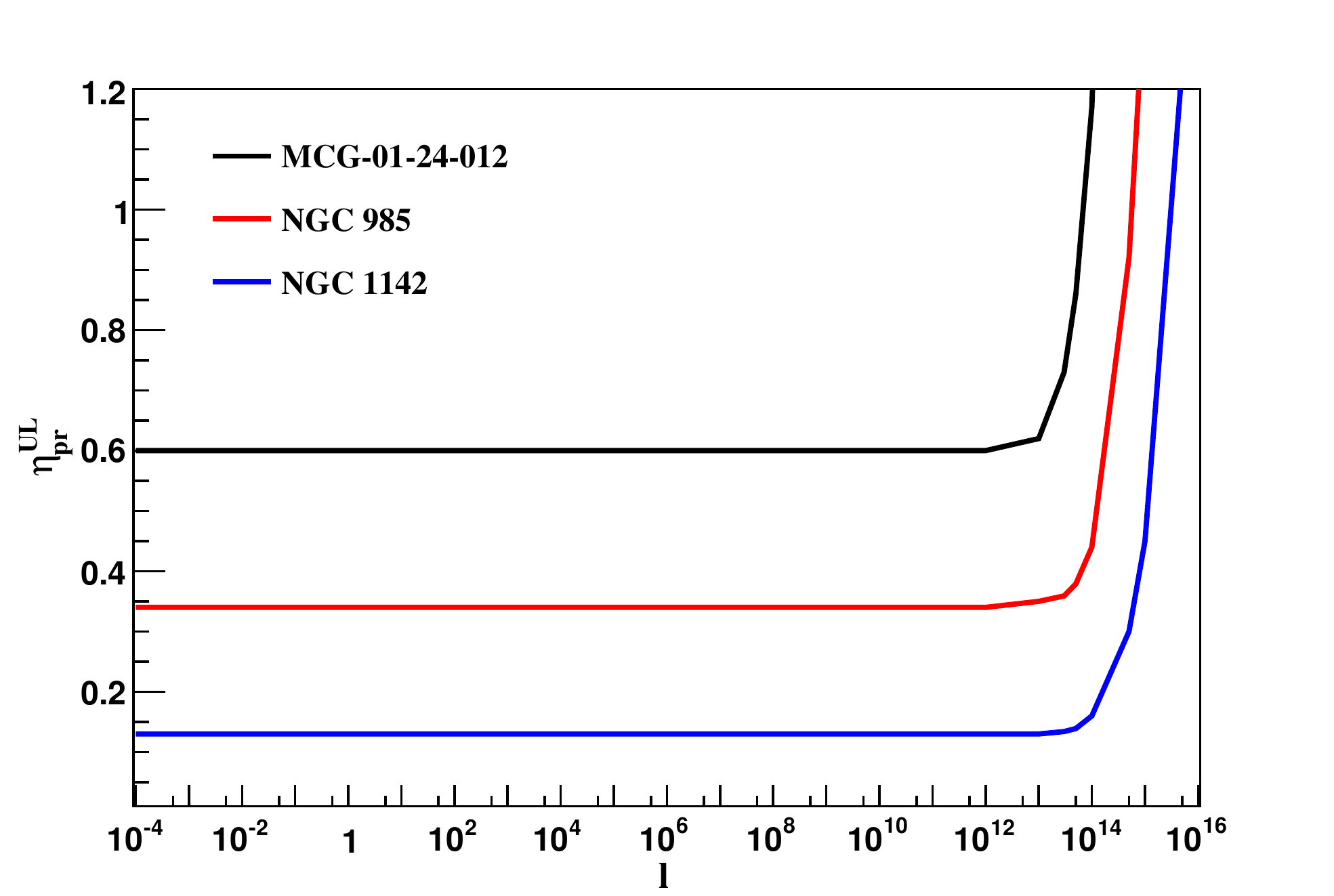}}
  \caption{The plots show the excluded regions with the values of $\ell$ to sources MCG-01-24-012,
    NGC 985 and NGC 1142. Figure (a) shows the variation of $L_{acc}^{Theory}$ as a
    function $\ell$. Figure (b) shows the variation of $\eta_{pr}^{UL}$
    as function of $\ell$.}
  \label{fig:luminosity:eta:l}
\end{figure}

Another important remark is that the theoretical bolometric luminosity $L_{acc}^{Theory}$ is sensitive to deviations of 
Newtonian gravity imposed by the presence of the AdS$_5$ bulk. This clearly can be seen in equation (\ref{dl}), since $R$ comes as 
solution of equation (\ref{cin}) and it is dependent of $\ell$
parameter, namely, the curvature radius of AdS$_5$ (sometimes called the 
warped compactication). Some values of $\ell$ promote huge gravity deviations and consequently the abrupt drop in luminosity shown in Figure \ref{fig:luminosity:eta:l}(a). To recover the upper limits of UHECR luminosities, smaller values of $L_{acc}^{Theory}$ give greater values of the conversion fraction $\eta_{pr}^{UL}$ (Fig. \ref{fig:luminosity:eta:l}(b)). The condition $\eta_{pr}^{UL}<1$ gives the allowed values of $\ell$ and consequently the allowed luminosity region. Since the excluded region only occurs when $\ell$ is really very large (and the existing limits point that $\ell$ should be $< 1$) then we can clearly adopt the $L_{acc}^{Theory}$ luminosities calculated in tables \ref{tab:1} and~\ref{tab:2}.        

In this paper we calculated the luminosity due to accretion of nine
point sources. These calculations lead to an parametrization of the
radius of a BH with extra dimension corrections as a function of its
mass. The luminosities calculated for nine specific source based on
theory were compared to an upper limit on the UHECR luminosity. The
comparison resulted in the determination of upper limits on the energy
conversion from accretion to UHECR. The calculations and theoretical
estimations of this conversion efficiency are rarely found in the literature and represent an important piece of information in the understanding of the energy balance in BH.

Given the upper limit on the proton or total UHECR luminosity used here, we did not manage to use these values to constrain extra dimension corrections. Such a limitation could be imposed if the proton or total UHECR luminosity were actually measured, this is currently beyond the experimental possibilities. Nevertheless, our studies show which sources could be more interesting to measure if this were the target. In principal one should look for the largest BH mass and the lowest upper limit on UHECR that could be set. Tables~\ref{tab:1} and~\ref{tab:2} summarize the findings of this paper.

\section{Acknowledgements}
RdR is grateful to CNPq (grants No. 303027/2012-6, No. 451682/2015-7 and No. 473326/2013-2), and to FAPESP (grant No.  2015/10270-0) for partial financial support. VdS thanks the support of the Brazilian population via CNPq and FAPESP (2014/19946-4). RCA and VdS thank the Pierre Auger Collaboration.

\begin{landscape}
\begin{table}[p]
\centering
\begin{tabular}{|c|c|ccc|c|cc|}
\hline
\textbf{Source name} & $\mathbf{log (M/M_{\odot})}$ & \multicolumn{3}{|c|}{$\mathbf{L_{acc}^{Theory}}$ [erg s$^{-1} \times 10^{45}$]}  & $\mathbf{L_{pr}^{UL}}$ & \multicolumn{2}{|c|}{$\mathbf{\eta_{pr}^{UL}}$} \\ 
 & & $R = R_{Sbrane}$ & $R = R_{\rm + Kerr}$ & $R = R_{\rm + Kerr brane}$ & [erg s$^{-1} \times 10^{45}$] & $R = R_{\rm + Kerr}$ & $R = R_{\rm + Kerr brane}$ \\ \hline

NGC 5995 & 9.89 & 1.16 & 6.23 & 0.58 & 0.90 & 0.14 & -  \\ \hline
2MASX J11454045-1827149 & 10.00 & 1.50 & 8.02 & 0.74 & 1.30 & 0.16 & - \\ \hline
MCG-01-24-012 & 10.16 & 2.17 & 1.16 & 1.08 & 0.65 & 0.56 &0.60  \\ \hline
Mrk 520 & 10.40 & 3.78 & 2.01 & 1.88 & 0.98 & 0.49 & 0.52  \\ \hline
2MASX J07595347+2323241 & 10.57 & 5.59 & 2.98 & 2.78 & 1.01 & 0.34 & 0.36 \\ \hline
LEDA 170194 & 10.59 & 5.85 & 3.12 & 2.91 & 1.48 & 0.47 & 0.50 \\ \hline
NGC 985 & 10.60 & 5.99 & 3.19 & 2.98 & 1.03 & 0.32 & 0.34 \\ \hline
CGCG 420-015 & 10.63 & 6.42 & 3.42 & 3.19 & 0.95 & 0.28 & 0.29  \\ \hline
NGC 1142 & 10.70 & 7.54 & 4.02 & 3.75 & 0.49 & 0.12 & 0.13 \\ \hline
\end{tabular}
\caption{Columns show the source name, the mass source according to
  references~\cite{lisa,michael}, the luminosity calculated with
  equation~\ref{dl} for $R = R_{{\rm Sbrane}}, R_{+\rm Kerr}$ or
  $R_{{+\rm Kerr brane}}$ with $\ell=10^{-4}\;m$, the upper limit on the proton UHECR luminosity according to equation~\ref{eq:CRUL} and the upper limits of $\eta_{pr}$ for $R = R_{+\rm Kerr}$ and $R_{{+\rm Kerr brane}}$.}
\label{tab:1}
\end{table}

\begin{table}[p]
\centering
\begin{tabular}{|c|c|ccc|c|cc|}
\hline
\textbf{Source name} & $\mathbf{log (M/M_{\odot})}$ & \multicolumn{3}{|c|}{$\mathbf{L_{acc}^{Theory}}$ [erg s$^{-1} \times 10^{45}$]}  & $\mathbf{L_{CR}^{UL}}$ & \multicolumn{2}{|c|}{$\mathbf{\eta_{CR}^{UL}}$} \\ 
 & & $R = R_{Sbrane}$ & $R = R_{\rm + Kerr}$ & $R = R_{\rm + Kerr brane}$ & [erg s$^{-1} \times 10^{45}$] & $R = R_{\rm + Kerr}$ & $R = R_{\rm + Kerr brane}$ \\ \hline
2MASX J11454045-1827149 & 10.0 & 1.50 & 8.02 & 0.74 & 3.16 & - & 0.23 \\ \hline
LEDA 170194 & 10.59 & 5.85 & 3.12 & 2.91 & 3.71 & - & 0.78 \\ \hline
NGC 985 & 10.6 & 5.99 & 3.19 & 2.98 & 2.19 & 0.69 & 0.73 \\ \hline

\end{tabular}
\caption{Columns show the source name, the mass source according to
  references~\cite{lisa,michael}, the luminosity calculated with
  equation~\ref{dl} for $R = R_{{\rm Sbrane}}, R_{+\rm Kerr}$ or
  $R_{{+\rm Kerr brane}}$ with $\ell=10^{-4}\;m$, the upper limit on the total UHECR luminosity according to equation~\ref{eq:CRUL} and the upper limits of $\eta_{CR}$ for $R = R_{+\rm Kerr}$ and $R_{{+\rm Kerr brane}}$.}
\label{tab:2}
\end{table}
\end{landscape}


\end{document}